\documentclass[12pt,amsfonts,amsmath]{article}
\usepackage{amssymb} 

\newtheorem{theorem}{Theorem}[section]
\newtheorem{definition}[theorem]{Definition}
\newtheorem{proposition}[theorem]{Proposition}
\newtheorem{lemma}[theorem]{Lemma}
\newtheorem{corollary}[theorem]{Corollary}

\renewcommand{\max}{^{\rm max}} 
\newcommand{\stat}{^{\rm stat}}\newcommand{\opp}{^{\rm opp}}
\newcommand{\eins}{\hbox{\rm 1}\mskip-4.4mu\hbox{\rm l}}
\newcommand{\End}{{\rm End}} \newcommand{\Tr}{{\rm Tr}\;}
\newcommand{\CZ}{{\mathcal Z}} \newcommand{\ZZ}{{\mathbb Z}} 
\newcommand{\QED}{\hspace*{\fill}Q.E.D.\vskip2mm}

\begin{document}

\title{Locality and Modular Invariance \\ in 2D Conformal
  QFT \footnote{Contributed to ``Mathematical Physics in Mathematics
    and Physics'', Siena, June 20-25, 2000}}

\author{K.-H. Rehren \\ Inst.\ f.\ Theor.\ Physik\\ Universit\"at G\"ottingen\\
  37073 G\"ottingen \\ Germany}
\date{\it Dedicated to S. Doplicher and J.E. Roberts.}
\maketitle

\begin{abstract}
The relations and differences between various classification problems 
arising in the context of local two-dimensional (2D) conformal quantum
field theory, modular invariants, and subfactors, are discussed. The
extent to which locality implies modular invariance, is exhibited. \\[1mm]
\end{abstract}

\section*{Introduction: \\ 
Modular invariants, 2D conformal QFT, and subfactors}
One of the great excitements in conformal QFT was the ADE classification
of modular invariant coupling matrices for the SU(2) current algebra
\cite{CIZ}. At each level $k$, this algebra has only a finite number
of covariant representations with positive energy (superselection
sectors), and the partition functions for the conformal Hamiltonian in
these sectors span a linear representation of the modular group
$SL(2,\ZZ)$ acting on the temperature parameter. A modular invariant
is a quadratic form (described by the coupling matrix $Z$) in this space of
partition functions which is invariant under the modular group, and
which is subject to a number of additional constraints. These
constraints are necessary for the modular invariant being
interpretable as the partition function of a local 2D conformal QFT 
with a unique vacuum vector. This 2D conformal QFT contains the SU(2)
current algebra both as left and right chiral subalgebras.    

The fact that $SU(2)$ modular invariants can be classified according to
an ADE scheme, has raised the interest of mathematicians both for its
arithmetic appeal, and for its relationship to several other
classification problems (for an overview see \cite{Z}). As a
physicist, I want to focus on the quantum field theoretical aspect of
this classification problem. I also want to consider it from a broader
perspective, namely as a statement about the possible position of the
embedding of two (left and right) chiral subtheories $A_L$ and $A_R$
into a local 2D conformal quantum field theory $B$. The case of
$SU(2)$ is then just a very special one, and an exact ADE
classification cannot be expected to prevail in general. 

From this perspective, the classification problem is to find all local 2D 
conformal QFT's which extend the given chiral subtheories $A_L$ and $A_R$. 
These extensions are also (though not completely) characterized by a
coupling matrix $Z$. The nontrivial requirement here which imposes
constraints on $Z$, is locality, while modular invariance is a
secondary feature.

Finally, there is a third aspect of modular invariants which is of
interest for the classification of subfactors. The embedding of chiral
subtheories into a local 2D conformal QFT gives rise to local
subfactors of the form $A\otimes C\subset B$ which are called
canonical tensor product subfactors \cite{CTPS}. The specific meaning
of ``canonical'' will be explained below, and is again related to the
existence of a coupling matrix. Such subfactors also arise, e.g., as
generalized quantum doubles or as asymptotic subfactors. A general
classification is not available, but several results on canonical
tensor product subfactors will be reported, which have direct implications 
with regard to the modular and local QFT classification problems.   

I emphasize that in the local perspective, modular invariance is an
additional requirement which is often imposed on the 2D conformal QFT
by string theory demands. In fact, it is not independent from the
requirement of locality. Several results below show that 2D locality
``almost'' implies modular invariance -- but not quite, since there
are easy examples of local 2D extension with a coupling matrix which
is not modular invariant. On the other hand, among the modular
invariant coupling matrices for affine Lie algebras of higher rank,
there are some accidental ones, which do not come from any local 2D
conformal QFT (e.g., \cite{SY}).    

In order to further clarify the relation between locality and modular
invariance, we recall the notion of ``statistics characters''
introduced in \cite{KHR} which directly derives from locality. It
provides a pair of matrices $X$ and $Y$ defined in terms of the
statistics of a system of superselection sectors (Sect.\ 3). $X$ is
the diagonal matrix of the statistics phases, while $Y$ collects the
values of the (relative) monodromy operators in a natural tracial
state. Unless there is some degeneracy in the statistics, these
matrices yield another unitary representation of the modular group
$SL(2,\ZZ)$ (``statistics representation'', Prop.\ 3.3). A priori,
this representation is not related to modular transformations of the
temperature parameter of the partition functions for these sectors. 
(Note also that partition functions need not even exist a priori as
Gibbs functionals.) But at least for affine Lie algebras and certain
related algebras, both representations of the modular group exist, and
coincide.    

The counter examples mentioned above (local 2D conformal QFT, but not modular
invariant) are due to the possible non-coincidence of the two
representations. In many cases, even if thermodynamic modular 
invariance fails, the analogous invariance property with respect to
the matrices $X$ and $Y$ still holds true. I propose to call this
invariance property ``(degenerate or nondegenerate) statistics
symmetry'' (Def.\ 3.4). Nondegenerate statistics symmetry is
equivalent to modular invariance with respect to the statistics 
representation (Prop.\ 3.5). Useful criteria for statistics symmetry
will be provided (e.g., the maximal chiral observables within the 2D
theory always have it), but there are also  examples of local 2D
theories which satisfy not even statistics symmetry, cf.\ Sect.\ 3. 

I shall not discuss the characterization of solutions to the various
classification problems in terms of graphs (such as ADE). Further
details on this aspect can be found in \cite{BE5,O,Z}. 

\section{The modular invariants perspective}\setcounter{equation}{0}

One considers a rational chiral QFT $A$ and the finite system of its
covariant representations $\pi_i$ with positive energy. The most
prominent examples for $A$ are affine Lie algebras (current algebras),
the Virasoro algebra with central charge $c<1$, or certain $W$-algebras 
arising in coset constructions. One assumes that the characters
$\chi_i(\beta)$ for each of these representations exist as Gibbs
functionals (i.e., $\exp -\beta \pi_i(L_0)$ are trace class operators), 
and that the characters transform linearly in a unitary representation
of the $SL(2,\ZZ)$ transformations of the complex temperature parameter 
$\tau=i\beta/2\pi$ in the regime ${\rm Im}(\tau)>0$ (and further parameters, 
if necessary). This assumption is fulfilled for all the examples mentioned 
\cite{KP}. 

The unitary matrix representatives for the generating transformations
$\tau\to\tau+1$ and $\tau\to -1/\tau$ are commonly called $T$ and $S$, 
respectively. The square of the latter transformation is a central
involution in $SL(2,\ZZ)$, and unity in $PSL(2,\ZZ)$. Hence $C=S^2$
commutes with $S$ and $T$, and $C^2=\eins$. The $SL(2,\ZZ)$ relations
are thus  
\begin{equation} 
  TSTST=S, \qquad CT=TC, \qquad CS=SC=S^{-1}.
\end{equation}
One looks for modular invariant quadratic forms of the form
\begin{equation} 
  \CZ(\tau)=\sum_{ij} Z_{ij}\;\chi_i(\tau)\chi_j(\overline\tau)
\end{equation} 
with a {\bf coupling matrix} $Z$, satisfying
\begin{equation} 
  TZ=ZT\quad\hbox{and}\quad SZ=ZS \qquad\hbox{(modular invariance)}. 
\end{equation}
In addition one postulates that the matrix entries $Z_{ij}$ are
nonnegative integers, and $Z_{00}=1$ (the label $0$ is always reserved
for the vacuum sector). These are necessary conditions if one wants to
interpret $\CZ$ as a partition function
$\Tr\exp(-{\rm Re}(\beta) P_{\rm conf}^0-i{\rm Im}(\beta)P_{\rm conf}^1)$
where $P_{\rm conf}^\mu$ are the conformal Hamiltonian and momentum of
a 2D conformal QFT with a unique vacuum vector. $Z_{ij}$ then is the
multiplicity of the product of chiral sectors $\pi_i\otimes \pi_j$
within the vacuum representation of the 2D theory. One considers   
\vskip4pt
  {\bf The modular classification problem:} 
  Find all coupling matrices $Z$ with nonnegative integer entries
  and $Z_{00}=1$, which commute with the given pair of matrices $S$ and
  $T$ (and find out and discard the accidental ones, which do not
  correspond to a local 2D theory, see above).
\vskip4pt
One finds several types of solutions, which come in pairs because with
$Z$ also $ZC=CZ$ is a solution. There is always the diagonal solution,
$\CZ=\sum_{i} \chi_i\otimes\chi_i$, along with the conjugate diagonal
solution $\CZ=\sum_{i} \chi_i\otimes\chi_{\bar i}$. All solutions
share the ``block form''  
\begin{equation} \CZ=\sum_{I} \chi_I\otimes\chi_{\sigma(I)} \end{equation} 
where the ``extended characters'' $\chi_I=\sum_{i} N_{iI}\chi_i$ unite
certain ``families'' $I$ of sectors, and $\sigma$ is some permutation
of the families. Among these, there are the orbifold solutions where
sectors $i$ and $j$ belong to the same family if and only if they differ
by a simple sector (with respect to fusion) belonging to some group of
simple sectors. Solutions with the identical permutation are called
Type I, those with a nontrivial permutation $\sigma$ are called Type II.  

For the $SU(2)$ affine Lie algebra, it turns out that for all modular
invariants, the nonvanishing diagonal entries $Z_{ii}\neq 0$ can be
identified with the Coxeter exponents of a Dynkin diagram of type A, D or E 
\cite{CIZ}. The $A_n$ diagrams correspond to the diagonal type I
invariants, $D_{\rm odd}$ to permutation and $D_{\rm even}$ to
orbifold invariants, respectively. $E_6$ and $E_8$ correspond to the 
exceptional Type I, and $E_7$ to the exceptional Type II invariants. 
These ADE invariants exhaust all modular invariants for $SU(2)$. There
are similar classifications for $SU(3)$, but an exact match with
Dynkin diagrams cannot be achieved \cite{G}.  

\section{The local QFT perspective}\setcounter{equation}{0}
We consider  
\vskip4pt
  {\bf The local classification problem:} 
  Find all local 2D conformal QFT's which irreducibly extend the given
  pair of chiral theories $A=A_L\otimes A_R$,
\begin{equation} A_L\otimes A_R\subset B. \end{equation}
\vskip4pt
To formalize the problem, I adopt the algebraic framework of QFT;
thus a QFT is given by a net of local von Neumann algebras (factors),
say $A(I)$ with $I$ a light ray interval, or $B(O)$ with $O$ a 2D
double-cone, subject to standard axioms including essential Haag
duality. An inclusion like (2.1) is thus always understood as the
collection of local subfactors $A_L(I)\otimes A_R(J)\subset B(O)$ for
all $O=I\times J$.   

There is no a priori reason from the local QFT perspective why the
left and right chiral subtheories $A_L$ and $A_R$ of a 2D conformal
QFT $B$ should be isomorphic (parity symmetric). I prefer to include the
``heterotic'' (non-symmetric) case in my discussion from the beginning.  

The left and right chiral observables within the 2D conformal theory
$B$ must commute with the respective opposite M\"obius group. It has
been shown \cite{CM} that all 2D observables which are invariant under
the right M\"obius group, indeed define a chiral subtheory, referred
to as the maximal left chiral observables $A_L\max$. These contain the
given subtheory of left chiral observables $A_L$, and one has the
intermediate inclusion  
\begin{equation} 
  A_L\otimes A_R\subset A_L\max\otimes A_R\max\subset B.
\end{equation}

Anticipating results below, one finds that the maximal chiral
observables coincide with the ``extended observables'' in the modular
invariants context \cite{MS}, with respect to which the coupling matrix
turns into a permutation matrix thus giving rise to the block form (1.4).
This fact is due, in view of Prop.\ 4.1 below, to the following
equivalent characterization of the maximal chiral observables \cite{CM}.

\begin{proposition} \cite{CM} 
  Let $B$ be a local 2D conformal QFT, and $A_L$, $A_R$ chiral
  subtheories whose observables generate the left and right M\"obius
  groups, respectively. For any double-cone region of conformal 2D
  Minkowski space, $O=I\times J$ in light ray coordinates, one has  
\begin{equation} A_L(I) \subset A_L\max(I) = B(O)\cap U(G_R)' = B(O)
  \cap A_R(J)'. 
\end{equation}
  (The corresponding staments hold for $R\leftrightarrow L$.) In
  particular, $A_L\max(I)$ and $A_R\max(J)$ are each other's relative
  commutants in $B(O)$.
\end{proposition}

The subalgebras of left and right chiral observables form an algebraic
tensor product, and at least in the cyclic subspace of the vacuum they
are also represented as a tensor product \cite{CM}. It is expected
that the tensor product is spatial in the full representation of the
2D theory, and we shall assume this in the sequel.

This means that the vacuum representation of $B$, as a representation
of the subalgebra $A_L\otimes A_R$ decomposes into irreducibles
according to the scheme
\begin{equation} 
  \pi\simeq\bigoplus_{ij}Z_{ij}\;\pi^L_i\otimes\pi^R_j\;. 
\end{equation} 
The coupling matrix $Z$ appearing in this decomposition serves as a
first (though not complete) characterization of the extension.
Again, $Z$ has nonnegative integer entries, and $Z_{00}=1$ because of
the uniqueness of the vacuum vector, but the labels $i$ and $j$ may
run over different sets (the sectors of $A_L$ and of $A_R$), and $Z$
may be rectangular. Although a priori the direct sum might be
countably or even uncountably infinite, we shall assume it to be
finite throughout. 

It follows that the thermodynamical partition function of the 2D 
conformal QFT, if it exists, is of the most general form in terms of
chiral Gibbs functionals,
\begin{equation}
  \CZ=\sum_{ij}Z_{ij}\;\chi^L_i(\tau)\chi^R_j(\overline\tau),
\end{equation} 
with the same coupling matrix $Z$ as in the decomposition (2.4). The
notions of ``coupling matrix'' thus coincide in the modular and in the 
local interpretation. I emphasize, however, that coupling matrices
associated with local 2D extensions are not necessarily modular
invariant. This issue will be discussed in Sect.\ 3.  

Due to the intermediate inclusion (2.2) the coupling matrix is a product 
\begin{equation} Z=B_L^tZ\max B_R \end{equation}
where $Z\max$ is the coupling matrix with respect to the maximal left
and right chiral observables, and the branching matrices $B_L$ and
$B_R$ describe the irreducible decomposition of the chiral superselection
sectors $\pi\max_I$ of $A\max$ upon restriction to $A\subset A\max$. The 
restricted representations $\pi\max_I\vert_A\simeq\bigoplus B_{Ii}\,\pi_i$ 
play a role like the families of sectors in eq.\ (1.4). This analogy
will also be exhibited in Sect.\ 3.

The question arises by which data of the chiral subtheories, apart
from the coupling matrix, their local 2D extensions can be characterized. 
(It should be stressed that the coupling matrix will in general not
determine the local extension up to equivalence, although it happens
to do so in models with ``few'' sectors.) An answer has been given in
\cite{LR} for the more general problem of characterizing local
extensions $B$ of a QFT $A$ in any dimension.  

The local extension $B$ of a quantum field theory $A$ is characterized
by its {\bf canonical DHR triple} $(\rho,w,w_1)$ (assuming the index
$\lambda$ of the local subfactors $A(O)\subset B(O)$ to be finite)
\cite{LR}. Here, $\rho$ is the DHR endomorphism \cite{DHR} of $A$
which describes the vacuum  representation of $B$ as a reducible
representation of $A$, and the isometric intertwiners
$w:\hbox{id}_A\to\rho$, $w_1:\rho\to\rho^2$ in $A$ satisfy the 
identities of a ``Q-system'' \cite{L} 
\begin{equation} 
  \begin{array}{c} w^*w_1=\rho(w^*)w_1=\lambda^{-\frac12} \eins_A, \\
  w_1w_1=\rho(w_1)w_1, \qquad w_1w_1^*=\rho(w_1^*)w_1, \end{array}
\end{equation} 
and the eigenvalue condition for the statistics operator \cite{DHR} 
$\varepsilon(\rho,\rho)\in\rho^2(A)'$ 
\begin{equation} \varepsilon(\rho,\rho) w_1=w_1. \end{equation}
The significance of the operator identities (2.7) and (2.8) has been
explained in \cite{RST}: The sector of $\rho$ describes the DHR charges
\cite{DHR} with respect to the subtheory $A$ which appear in the
vacuum representation of $B$. The isometry $w$ singles out the vacuum
(charge zero) sector within $\rho$, and $w_1$ is a generating functional 
for the collection of $3j$-symbols (amplitudes of 3-point functions of
charge carrying fields). The identities (2.7) reflect the condition
that the corresponding linear combinations of localized vertex
operators (charged field bundle elements \cite{FRS}) form a * algebra. 
The eigenvalue condition (2.8) expresses the condition that these
combinations, which constitute the fields of $B$, commute at spacelike
distance.   

The extension $B$ is recovered from the canonical DHR triple as
follows \cite{LR}. If $\rho$ is localized in some space-time
double-cone $O_0$, then it restricts to an endomorphism of the
corresponding local algebra, $\rho_0\in\End(A(O_0))$, and 
$(\rho_0,w,w_1)$ is a Q-system which determines the local algebra
$B(O_0)$ extending $A(O_0)$. With the help of unitary charge
transporters for $\rho$, the local Q-system $(\rho_0,w,w_1)$ can be
moved to every Poincar\'e or M\"obius transformed double-cone
$O=g(O_0)$, giving rise to a coherent net of algebras $B(O)$ extending
$A(O)$. These satisfy local commutativity thanks to (2.8).

We conclude that local extensions of local QFT's are completely
characterized by canonical DHR triples, that is, solutions $(\rho,w,w_1)$ 
to (2.7) and (2.8) within the category of DHR endomorphisms and their
intertwiners. For rational QFT's, these operator identities constitute
a finite system of nonlinear algebraic equations which may serve as
the basis of the present classification problem. 

In the 2D case of our interest, $A=A_L\otimes A_R$ is a tensor
product, and so is its DHR category. The endomorphism $\rho$ entering
the problem is given by eq.\ (2.4),
\begin{equation}
  \rho\simeq\bigoplus_{ij}Z_{ij}\;\alpha^L_i\otimes\alpha^R_j.
\end{equation} 
Its specification is equivalent to the specification of the coupling
matrix. Thus, the local classification problem is equivalent to 
\vskip4pt
  {\bf The local classification problem (algebraic version):} 
  Find all canonical DHR triples for the given pair of chiral
  theories $A=A_L\otimes A_R$ with $\rho$ of the form (2.9).
\vskip4pt
The local classification problem involves the determination of all
admissible coupling matrices. As was mentioned before and will be
discussed in more detail in the next section, these do not coincide
with the modular invariant coupling matrices. Nevertheless, the
question arises whether for any given modular invariant, an associated 
local 2D conformal QFT exists, and whether this will be unique. We
therefore address also 
\vskip4pt
  {\bf The existence and uniqueness problem for local 2D conformal QFT:} 
  Decide whether a given coupling matrix $Z$ for $A_L\otimes A_R$
  (modular invariant or not) arises as the coupling matrix of a local
  2D conformal QFT, and whether the latter is unique.\vskip4pt

There is no reason to expect that the uniqueness problem will 
have a positive answer in general. It is well known that
non-isomorphic subfactors can have the same canonical endomorphism
(hence the same coupling matrix). The eigenvalue condition (2.8) will
presumably not alter the situation very much. 

As for the existence problem, some progress has been made. A standard
solution of (2.7) and (2.8) has been given in \cite{LR} for the
coupling matrix $Z=C$ pertaining to the parity symmetric case
$A_L\simeq A_R$, based on Prop.\ 4.2 below. This means that conjugate
left and right chiral DHR charges can always be combined to yield
local 2D fields. The standard solution exists for any closed (under
fusion and conjugation) subsystem of DHR sectors of $A$, and thus
yields truncated conjugate diagonal coupling matrices which are
typically not modular invariant.    

A more general class of possibly heterotic solutions can be obtained
from Prop.\ 4.3 below for all $SU(2)$ and many other modular invariant
coupling matrices. 

\section{On the relation between modular invariance and locality}
\setcounter{equation}{0}
It is appropriate to discuss the distinction between the local and the
modular classification problem. 

First, it is not completely understood in which sense the Gibbs
partition functions of any chiral theory $A$ should transform linearly
under the modular group in general when no Kac-Peterson formula
\cite{KP} is available. (Higher rank current algebras already exhibit
pairs of conjugate sectors with the same partition function, and this
degeneracy has to be lifted by introducing additional thermodynamic
parameters on which the modular group can act.) The most general
result in this direction is due to Nahm \cite{N}. Only the subgroup
generated by $T$ acts in general, which measures the spectrum of $L_0$
modulo integers.    

Second, one should not exclude heterotic models with different left and
right chiral observables, as such models might arise upon passage to
``extended chiral algebras'' \cite{MS}, even if one starts with
parity symmetric models. The left and right chiral algebras then might
have different modular transformation matrices, and the requirement of
modular invariance is no longer a commutation property, but an
intertwining property:
\begin{equation} 
  T_LZ=ZT_R\quad\hbox{and}\quad S_LZ=ZS_R
  \qquad\hbox{(heterotic modular invariance)}. 
\end{equation}
If we enlarge the framework accordingly, the question arises to which
extent modular invariance of the coupling matrix of a local 2D
extension is implied by the properties of a canonical DHR triple. 

The intertwining property for $T$ is easily established. According to
the standard argument, all Wightman fields affiliated with the local
2D theory $B$ must be Bose fields, hence should have integer
difference between left and right chiral scaling dimensions.  
In the algebraic framework, this follows from the eigenvalue condition: 

\begin{lemma} 
  Let $(\rho,w,w_1)$ be a canonical DHR triple (in any dimension), and
  $\sigma$ an irreducible subsector of $\rho$. Then $\sigma$ has
  statistics phase $+1$. 
\end{lemma}
\par\noindent {\it Proof:}  
The statistics parameter of $\rho$ \cite{DHR}
$$ \phi_\rho(\varepsilon(\rho,\rho))=r^*\rho(\varepsilon(\rho,\rho))r=
\rho(r^*)\varepsilon(\rho,\rho)^*r $$
can be computed: inserting the isometry $r=w_1w:\hbox{id}_A\to\rho^2$,
and using (2.8) and (2.7) yields
$$ \phi_\rho(\varepsilon(\rho,\rho)) = \rho(w^*w_1^*)w_1w =
\rho(w^*)w_1w_1^*w = \lambda^{-1}\eins=d(\rho)^{-1}\eins. $$
Since the spectrum of the statistics parameter determines the
statistics phases of the subsectors \cite{DHR,FRS}, the claim follows. 
\QED

\begin{corollary} In the 2D conformal case, with $A=A_L\otimes A_R$,
  let $\sigma\prec\rho$ be of the tensor product form
  $\sigma_L\otimes\sigma_R$. Then $\sigma_L$ and $\sigma_R$ have equal
  statistics phases. Hence the coupling matrix intertwines the left
  and right diagonal matrices $X$ of statistics phases,
\begin{equation} X_L Z=Z X_R. \end{equation}
\end{corollary}
\par\noindent {\it Proof:}  Spacelike separation in 2D means positive left 
separation and negative right separation, or vice versa. Hence the 
statistics operator of $\sigma$ is the tensor product of two opposite 
chiral statistics operators, and the statistics phase of $\sigma$ is
the quotient of the chiral statistics phases. From this, the
statements are obvious.  
\QED 

Now, the chiral statistics phases $\kappa$ are related to the chiral
scaling dimensions $h$ by the spin-statistics theorem \cite{GL},
$\kappa=\exp 2\pi i h$. Thus, the matrices $T$ differ from the matrices
$X$ by an overall complex phase $\exp 2\pi i\frac c{24}$ depending on
the chiral central charges \cite{KP}. It follows that the coupling
matrix $Z$ intertwines $T_L$ and $T_R$ up to the quotient of these
phases. Thus, locality always implies invariance under the modular
transformation $T$ up to a phase which is trivial if the left and
right central charges coincide modulo 24.  

A statement of comparable generality cannot be obtained for the full
modular group. E.g., the local 2D theory $B=A_L\otimes A_R$ has the
coupling matrix $Z_{ij}=\delta_{i0}\delta_{j0}$ which is definitely
not modular invariant. But we can obtain nontrivial general results 
concerning a symmetry which specializes to modular invariance in
favorable cases. As this symmetry does not refer to Gibbs functionals but
to statistics, I call it ``statistics symmetry'' (Def.\ 3.4). We
shall see that it is always satisfied, as a consequence of locality,
for the maximal chiral observables.  

Thus, rather than the thermodynamical transformation matrices $T$ and
$S$, we consider the statistics phase matrix $X$ (the same as in
Cor.\ 3.2) and the statistics character matrix $Y$, defined in
\cite{KHR}. These matrices can be written as     
\begin{equation} 
  X_{ij}=X_{ji}=\kappa_i\;\delta_{ij} \qquad\hbox{and}\qquad  
  Y_{ij}=Y_{ji}=\sum_{k} N_{ij}^k\;\frac{\kappa_i\kappa_j}{\kappa_k}\;d_k 
\end{equation}
where $i,j,k$ run over all irreducible DHR sectors of the model (in
the present case: of the chiral observables $A_L$ or $A_R$), or over
any closed subsystem $\Delta$ thereof. $\kappa_i$ and $d_i$ are the
complex phase and the inverse modulus of the statistics parameter
\cite{DHR,FRS} of the sector $\pi_i$, and $N_{ij}^k$ are the fusion
rules. For a non-exhaustive system $\Delta$, the matrices $X$ and $Y$
are truncated to those rows and columns of the full matrices which
correspond to $\Delta$.  

\begin{proposition} \cite{KHR} 
  The matrices $X$ and $Y$ satisfy the relations
\begin{equation} 
  XYXYX=z\;Y, \qquad CX=XC, \qquad CY=YC=Y^*
\end{equation}
  where $z=\sum_{i\in\Delta}\kappa_id_i^2$ and $C$ is the 
  charge conjugation matrix. If $Y$ is invertible, a rescaling
  of $X$ and $Y$ yields unitary matrices $T\stat$ and $S\stat$ which
  satisfy (1.1) and hence generate a representation (``statistics
  representation'') of $SL(2,\ZZ)$, turning $\Delta$ into a modular
  category \cite{MM}. 

  (More specifically, in the nondegenerate case, $\vert z\vert^2$ equals 
  the ``global index'' $w=\sum_{i\in\Delta}d_i^2$ of the system of
  sectors, and $S\stat=w^{-1/2}Y$, $T\stat=(z/\vert z\vert)^{-1/3}X$). 
\end{proposition}

It is expected that $T\stat=T$ and $S\stat=S$, whenever modular
transformation matrices $T$ and $S$ exist independently. In fact, the
first of these coincidences is essentially the spin-statistics theorem
\cite{GL} which relates the statistics phase determining $T\stat$ to
the chiral scaling dimension (``spin'') entering $T$. The second one
is at least empirically true wherever it has been tested.

That $Y$ is indeed invertible provided $\Delta$ is the system of {\em all} 
DHR sectors, was recently shown \cite{KLM} to follow from the split
property, and is therefore true whenever the chiral Gibbs functionals
$\Tr\exp-\beta\pi_i(L_0)$ exist for any temperature $1/\beta>0$.  

Let us now turn to the question of statistics symmetry. 

\begin{definition} A local 2D extension 
  satisfies {\bf statistics symmetry} if 
\begin{equation} 
  X_LZ=ZX_R\quad\hbox{and}\quad\frac1{\lambda_L}Y_LZ=\frac1{\lambda_R}ZY_R
  \qquad\hbox{(statistics symmetry)}.
\end{equation}
\end{definition}

The factors $\lambda_L=\sum_i Z_{i0}\, d(\pi^L_i)$ (and likewise
$\lambda_R$) in the $Y$ relation are dictated by comparison of the
$00$ components. They equal the indices $\lambda=d(\pi_0\max\vert_{A})$ of
the inclusions $A\subset A\max$ (because $(B_R)_{J0}=\delta_{J0}$ and
$Z\max_{I0}=\delta_{I0}$, hence $Z_{i0}=(B_L)_{0i}$).

\begin{proposition} If $Y_L$ and $Y_R$ are nondegenerate, statistics symmetry
implies 
\begin{equation} 
  T_L\stat Z=ZT_R\stat\quad\hbox{and}\quad S_L\stat Z=ZS_R\stat
  \quad\hbox{(nondegenerate statistics symmetry)},
\end{equation}
  i.e., the coupling matrix is a modular invariant with respect to the
  left and right statistics representations of $SL(2,\ZZ)$ (cf.\ Prop.\ 3.3). 
\end{proposition}
\par\noindent {\it Proof:}  The claim is clearly true up to scalar factors. Eqs.\
(3.4) and (3.5) imply $z_L/\lambda_L=z_R/\lambda_R$. The complex
phase and the modulus of this equality imply that all scalar factors 
cancel. (Incidentally, \cite[Prop.\ 3.1]{BEK} implies that the left
and right global indices $w_L=w_R$ and the inclusion indices
$\lambda_L=\lambda_R$ coincide separately.)
\QED

Cor.\ 3.2 states that the $X$ part of statistics symmetry follows
directly from the locality condition on the 2D extension. Locality
also enters the proof of the $Y$ part, but suitable completeness
conditions will be required in addition, cf.\ Cor.\ 3.8.

In order to discuss the intertwining relation for $Y$, we recall from
Sect.\ 2 the product form of the total coupling matrix
$$ Z=B_L^tZ\max B_R, \eqno(2.6)$$
due to the intermediate inclusion
$$  A_L\otimes A_R\subset A_L\max\otimes A_R\max\subset B. \eqno(2.2)$$
Here $Z\max$ is the coupling matrix for the maximal observables, and
$B_L$ and $B_R$ are the branching matrices. From Prop.\ 2.1 we know
that the maximal chiral observables are each other's relative
commutants, and in fact are distingushed by this property. Prop.\ 4.1
below will tell us that the sets of sectors of $A_L\max$ and
$A_R\max$ contributing to the coupling matrix (2.6) are both closed
under fusion, and $Z\max$ is a permutation matrix which induces an
isomorphism of the fusion rules, and hence preserves the statistical
dimensions. Since it also respects the statistics phases (Cor.\ 3.2),
it intertwines the left and right statistics characters $Y\max$ as
well. In particular, $Y_L\max$ is invertible if $Y_R\max$ is, and the
coupling matrix intertwines the ensuing left and right statistics
representations. 

\begin{corollary}
  The coupling matrix of a local 2D conformal QFT with respect to its
  maximal chiral observables (or equivalently: every coupling matrix which
  is a permutation matrix) satisfies 
\begin{equation} Y_L\max Z\max=Z\max Y_R\max, \end{equation}
  hence statistics symmetry. If in addition both the left and right
  chiral statistics is nondegenerate, then the coupling matrix is a
  modular invariant with respect to the left and right statistics
  representations of $SL(2,\ZZ)$. 
\end{corollary} 

In \cite{BE4} conditions were found that the branching matrices 
intertwine the matrices $Y$ and $Y\max$ up to a numerical factor
This implies that the total coupling matrix $Z=B_L^tZ\max B_R$ also
interwines $Y_L$ with $Y_R$, i.e., $Z$ has statistics symmetry:

\begin{proposition} \cite[Lemma 6.3]{BE4} 
  Consider a local extension (of chiral QFT's) $A\subset A\max$ with
  finite index $\lambda$. If $\Delta$ is a closed system of DHR
  sectors of $A$, then its image $\alpha^0(\Delta)$ under ambichiral
  induction is a closed system of DHR sectors of $A\max$, and the
  branching matrix $B$ intertwines the matrices $Y$ and $Y\max$
  associated with the systems $\Delta$ and $\alpha^0(\Delta)$: 
\begin{equation} \frac1\lambda BY= Y\max B. \end{equation}
\end{proposition}

``Ambichiral induction'' is a map between closed systems of DHR 
endomorphisms, based on $\alpha$-induction. The latter is the natural 
prescription to extend DHR endomorphisms of a subtheory to an ambient
theory, giving rise to soliton-type sectors in general. 
$\alpha$-induction was first considered by J. Roberts \cite{JR} in a 
cohomological problem, and its relevance for the local extension
problem was discussed in \cite{LR}. Its functorial properties where
first elaborated by Xu \cite{X} and by B\"ockenhauer and Evans \cite{BE}. 
Due to nontrivial monodromies, there are in fact two prescriptions for
$\alpha$-induction, and ambichiral induction is the intersection of
the two sets of irreducible subsectors obtained by applying both
prescriptions to the elements of $\Delta$.   

An entry of the branching matrix $B$ is by definition the multiplicity
of an irreducible DHR sector of $A$ in the restriction of an irreducible 
DHR sector of $A\max$. By ``$\alpha$-$\sigma$-reciprocity'' which
holds for local extensions \cite{BE}, this multiplicity equals the
multiplicity of the DHR sector of $A\max$ in the image of the sector
of $A$ under either $\alpha$-induction. In particular, every DHR
sector of $A\max$ belongs to the ambichiral image of each sector of
$A$ contained in its restriction. 

\begin{corollary} 
  Consider the local 2D extension $B$ with coupling matrices $Z$ and
  $Z\max$ with respect to the chiral observables $A_L \otimes A_R$ and
  $A_L\max \otimes A_R\max$, respectively. Let $\Delta_X$ and
  $\Delta_X\max$ ($X=L,R$) be the sets of chiral DHR sectors
  contributing to the respective coupling matrices. If $\Delta_X$ are
  both closed systems, and $\Delta_X\max$ both coincide with the
  images of $\Delta_X$ under ambichiral induction (or, equivalently,
  with the preimage of $\Delta_X$ under restriction), then the
  coupling matrix $Z$ satisfies statistics symmetry. 
\end{corollary}
\par\noindent {\it Proof:} 
The equivalence of the two conditions on $\Delta_X\max$ follows from
$\alpha$-$\sigma$-reciproc\-ity and the obvious fact that $\Delta_X$ is
the image of $\Delta_X\max$ under restriction. According to Prop.\
3.7, the branching matrices intertwine the statistics characters for
the closed systems of sectors $\Delta_X$ and $\alpha^0(\Delta_X)$,
while according to Cor.\ 3.6, the permutation matrix $Z\max$
intertwines the statistics characters for the closed systems
$\Delta_L\max$ and $\Delta_R\max$. Thus, under the stated conditions,
Cor.\ 3.6 and Prop.\ 3.7 show that the coupling matrix $Z=B_L^tZ\max B_R$
intertwines the statistics characters for $\Delta_L$ and $\Delta_R$
with the stated numerical factors.  
\QED

This corollary is the strongest result one may expect for general
heterotic theories, as a generalization of modular invariance 
implied by 2D locality. However, the conditions on the behavior of
sectors under induction and  restriction in the corollary (in order
to apply the results of \cite{BE4}) are not always satisfied. 

Simple counter examples of perfectly local 2D extensions which
violate the assumption and conclusion of Cor.\ 3.8 are given by
\begin{equation} 
  A_L\otimes A_R\subset A_L\max\otimes A_R\max = B
\end{equation} 
where the chiral inclusions $A\subset A\max$ have subfactor depth
larger than 2, such as the ``conformal embedding'' 
$SU(2)_{10}\subset Sp(5)_1$ with partition function  
\begin{equation} 
  \CZ=\chi_0\max\otimes\chi_0\max=(\chi_0+\chi_3)\otimes(\chi_0+\chi_3).
\end{equation}
Only those sectors of $A$ contribute to the coupling matrix which are
contained in the vacuum sector of $A\max$, and these do not form a
closed system unless the depth is 2. $\alpha$-induction on these
sectors produces new DHR sectors of $A\max$ which are not contained in
the vacuum representation of $B$. Completing the systems of chiral
sectors ensures the correct intertwining property for the branching
matrices, but the coupling matrix $Z\max$, now being a bijection
between {\em subsystems} of the completed systems, no longer
intertwines $Y_L\max$ with $Y_R\max$ for the {\em completed} systems.  

I do not see, whether the coupling matrix satisfies any sensible
intertwining property weaker than statistics symmetry in complete
generality. A possible criterium to exclude models like the counter
examples (3.10), and hopefully to enforce the intertwining property 
(3.5), could be that the local 2D theory $B$ does not possess
nontrivial superselection sectors, but I have no proof that this
condition indeed has the desired consequences. 

Thus, we see that 2D locality comes close to imply the statistics
symmetry (Def.\ 3.4) which generalizes modular invariance to general
heterotic models and to models without proper modular transformation
laws for the chiral Gibbs functionals. But counter examples are easily
constructed.  

Conversely, as mentioned in the introduction, there are accidental modular
invariants which do not admit a corresponding local 2D extension of
the chiral observables. Thus, we conclude that modular invariance and
2D locality are intimately related while neither can imply the other
without suitable further input.

\section{Canonical tensor product subfactors}\setcounter{equation}{0}

Solving the operator identities (2.7) within the DHR category of the
local net $A$ is equivalent to finding a Q-system $(\rho_0,w,w_1)$ for
the local factor $A(O_0)$ such that $\rho_0\in\End_{\rm fin}(A(O_0))$
is the restriction of some DHR endomorphism of $A$ localized in $O_0$. 
The problem is thus of the type: Find all subfactors $A_1$ of a factor
$A$ whose canonical endomorphisms $\rho$ decompose into irreducibles
from a given system of sectors of $A$. In this form it is
mathematically well-posed with any purely infinite factor $A$ and 
any given system of sectors $\Delta\subset \End_{\rm fin}(A)$. To
impose also the eigenvalue condition (2.8), one has to assume the
system of sectors to be braided.   

There is no general solution to this problem. But there are some
partial results which relate to the specific tensor product
structure. We call a subfactor of the form $A\otimes C\subset B$ a
{\bf canonical tensor product subfactor} (CTPS) \cite{CTPS} if its
dual canonical endomorphism $\rho\in\End_{\rm fin}(A\otimes C)$
decomposes into irreducibles as  
\begin{equation} 
  \rho\simeq\bigoplus_{ij} Z_{ij}\;\alpha_i\otimes\gamma_j,
\end{equation}
with $\alpha_i\in\End(A)$ and $\gamma_j\in\End(C)$.

The embeddings $A_L\otimes A_R\subset B$ of Sect.\ 3 share this
structure. The coupling matrix $Z$ of a CTPS is again in general
rectangular. It has nonnegative integer entries, and $Z_{00}=1$ if and
only if the CTPS is irreducible. We shall consider 
\vskip4pt
  {\bf The subfactor classification problem:} 
  Find all irreducible canonical tensor product subfactors 
  $A\otimes C\subset B$ with $\alpha_i$ and $\gamma_j$ in (4.1) belonging
  to given systems of sectors of the given factors $A$ and $C$, respectively.
\vskip4pt
We have seen in Sect.\ 2 that the left and right maximal chiral
observables are each other's relative commutants within the 2D
theory. Therefore, the following general result, formulated in the
broader framework of subfactors, is of interest.

\begin{proposition} \cite{CM} 
  Let $A\otimes C\subset B$ be a CTPS with dual canonical endomorphism
  as in (4.1). Then the following are equivalent.

(i) $\eins\otimes C$ is the relative commutant of $A\otimes\eins$ in
  $B$, and vice versa. (This property is called ``normality'').

(ii) $Z_{0j}=\delta_{0j}$ and $Z_{i0}=\delta_{i0}$.

(iii) The sectors $\alpha_i$ of $A$ and the sectors $\gamma_j$ of $C$
  contributing to (4.1) are both closed under conjugation and
  fusion. There is a bijection between them which preserves the fusion
  rules, and the coupling matrix is the permutation matrix
  for this bijection.
\end{proposition}

It is not clear whether every CTPS $A\otimes C\subset B$ has
an intermediate normal CTPS $\hat A \otimes \hat C\subset B$ as in Prop.\
4.1. Presumably, this is not the case in general. The results in
\cite{CM} on the tensor product position of chiral observables within
a 2D conformal QFT, however, ensure the existence of a unique normal
intermediate subfactor, corresponding to the maximal observables, 
$$  A_L\otimes A_R\subset A_L\max\otimes A_R\max \subset B. \eqno(2.2)$$

The proposition, when applied to canonical DHR triples, thus splits
the local classification problem into two independent parts:
classification of {\em chiral} local extensions $A\subset A\max$, and
classification of {\em normal} canonical DHR triples with $Z\max$ an
isomorphism of the fusion rules of $A_L\max$ and $A_R\max$.

The proposition also implies that the coupling matrix (2.6) for a local
2D extension is of block form (1.4) as for modular invariants. (Note
that in the heterotic case a distinction between ``Type I'' and ``Type
II'' is meaningless.)  

The CTPS classification problem can again be formulated algebraically in
terms of Q-systems for $A\otimes C$ satisfying (2.7) with $\rho$ of the 
form (4.1). We cannot solve this problem systematically, but there are
systematic prescriptions for the construction of Q-systems for
CTPS's. In this way, the following Props.\ 4.2 and 4.3 were obtained,
which are mainly relevant for the associated existence problem. 

\begin{proposition} \cite{LR} 
  Let $A$ be a purely infinite factor, and 
  $\Delta\subset\End_{\rm fin}(A)$ a finite closed system of irreducible
  endomorphisms $\sigma$ with finite index. Then there is a subfactor
  $A\otimes A\opp\subset B$ with dual canonical endomorphism
\begin{equation}
  \rho\simeq\bigoplus_{\sigma\in\Delta} \sigma\otimes\sigma\opp.
\end{equation}
  If the system $\Delta$ is braided, then the eigenvalue condition (2.8) is
  also satisfied.
\end{proposition}

The result has been generalized to factors of type II and to infinite
systems $\Delta$ by Masuda \cite{M}, who also showed that the resulting
subfactor is isomorphic to the asymptotic subfactor associated with
the inclusion $\rho_\Delta(A)\subset A$ where $\rho_\Delta$ is the
direct sum of all $\alpha_i\in \Delta$. Izumi has shown how to compute
the complete structure of these subfactors \cite{I}. 

The coupling matrix in Prop.\ 4.2 is the diagonal unit matrix for the
system $\Delta$. In chiral conformal QFT on the circle, $A\opp=A(I)\opp$ is
naturally isomorphic to $A(I)'=A(I')$, and the latter is M\"obius
conjugate to $A(I)=A$. Under these identifications, the opposite 
$\sigma\opp$ of a DHR endomorphism $\sigma$ belongs to the charge
conjugate sector $\bar\sigma$ \cite{GL}. Hence, in CQFT eq.\ (4.2)
corresponds to a canonical DHR triple for $A_L\simeq A_R$ with
coupling matrix $Z=C$, the charge conjugation matrix.  

Applied to the existence problem for a given coupling matrix, or
modular invariant, the proposition thus states that the conjugate
diagonal matrix $Z=C$ always corresponds to a local 2D theory. By
choosing irreducible subfactors $A_\nu \subset A$ ($\nu=1,2$), one obtains
an abundance of CTPS's $A_1\otimes A_2\opp\subset B$ with coupling
matrix in block form as with Type I modular invariants. The
corresponding families of sectors of $A_\nu$ (cf.\ Sect.\ 1) are the
restrictions of $\sigma\in\End_{\rm DHR}(A)$ to $A_\nu$.  

The following result is an important generalization, by which local 2D
conformal QFT's can be associated with numerous other (also Type II) modular
invariants. 

\begin{proposition} \cite{CTPS} 
  Let $A_\nu\subset M$ ($\nu=1,2$) be two subfactors of finite index,
  $\Delta_\nu\subset\End_{\rm fin}(A_\nu)$ two finite closed systems of
  (inequivalent irreducible) endomorphisms. For a pair of inductions
  $\widehat\cdot$ (see below) of $\Delta_\nu$ put    
\begin{equation} 
  Z_{\sigma,\tau} = \dim Hom(\widehat\sigma,\widehat\tau)_M
  \qquad(\sigma\in\Delta_1,\tau\in\Delta_2).
\end{equation}
  Then there is a CTPS $A_1\otimes A_2\opp\subset B$ with dual
  canonical endomorphism 
\begin{equation} 
  \rho\simeq\bigoplus_{\sigma\in\Delta_1,\tau\in\Delta_2}
  Z_{\sigma,\tau}\;\sigma\otimes\tau\opp.
\end{equation}
  (An ``induction'' $\widehat\cdot$ assigns to each $\sigma\in\Delta$ an
  endomorphism $\widehat\sigma\in\End_{\rm fin}(M)$ in a functorial manner: 
  $\widehat\sigma$ extends $\sigma$, i.e., $\widehat\sigma\vert_A=\sigma$, 
  and intertwiners between products of $\sigma_i$ are intertwiners for
  products of $\widehat\sigma_i$ as well, i.e.,
  $Hom(\sigma_1,\sigma_2\sigma_3)_A\subset
  Hom(\widehat\sigma_1,\widehat\sigma_2\widehat\sigma_3)_M$.)   

  If furthermore the systems $\Delta_\nu$ are braided, and the braidings
  fulfill a naturality condition with respect to
  $Hom(\widehat\sigma,\widehat\tau)_M$, then the CTPS also satifies 
  the eigenvalue condition (2.8).
\end{proposition}

Let $A_1 \simeq A_2$ be local algebras of chiral observables,
$\Delta_1 \simeq \Delta_2$ subsystems of their DHR endomorphisms, and 
assume that the dual canonical endomorphisms for the subfactors
$A_\nu\subset M$ are also DHR endomorphisms (these subfactors need 
{\em not} describe local chiral extensions). Then $\alpha$-induction
(cf.\ Sect.\ 3) is applicable to $\Delta_\nu$, and the assumptions of
Prop.\ 4.3 are fulfilled with the two opposite prescriptions for
$\alpha$-induction. It was shown in \cite{BEK} that the corresponding
coupling matrix (4.3) has statistics symmetry. In fact, many
thermodynamical modular invariants can be obtained this way. For
these modular invariants, Prop.\ 4.3 thus proves the existence 
of an associated local 2D conformal QFT. 

The statement of Prop.\ 4.3 is, however, not restricted to 
$\alpha$-induction, and therefore produces a larger class of CTPS's
than appearing in 2D conformal QFT. On the other hand, uniqueness of
the subfactors in Props.\ 4.2 and 4.3 is not claimed. There may well be 
inequivalent CTPS's with the same dual canonical endomorphism, but
with inequivalent Q-systems.

\section*{Conclusions}
I have discussed the interrelations between several classification
problems arising in mathematics and physics with various applications
and interpretations: modular invariants, local 2D conformal QFT, and
canonical tensor product subfactors. They all have some aspects in
common (notably a coupling matrix), but the specific requirements
imposed on the coupling matrix depend on the perspective. Progress in
any of these classification problems will have a bearing on the
related ones.

\section*{Acknowledgments}
I thank J. B\"ockenhauer, S. Carpi and A. Ocneanu for numerous
discussions, and J. Mund for a critical reading of the manuscript.

\end{document}